\documentstyle[12pt]{article}

\begin{document}

\title{
Defending  Time-Symmetrized Quantum Counterfactuals}
\author{ Lev Vaidman}
\date{}
\maketitle

\begin{center}
{\small \em School of Physics and Astronomy \\
Raymond and Beverly Sackler Faculty of Exact Sciences \\
Tel Aviv University, Tel-Aviv 69978, Israel. \\}
\end{center}

\vspace{2cm}
\begin{abstract}
  Recently, several authors have criticized the time-symmetrized
  quantum theory originated by the work of Aharonov et al. (1964).
  The core of this criticism was a proof, appearing in various forms,
  which showed that the counterfactual interpretation of
  time-symmetrized quantum theory cannot be reconciled with standard
  quantum theory.  I, (Vaidman, 1996a, 1997) have argued that the apparent
  contradiction is due to a logical error and introduced  consistent 
  time-symmetrized quantum counterfactuals. Here I  repeat my arguments
  defending the time-symmetrized quantum theory and reply to the
  criticism of these arguments by Kastner (1999).
\end{abstract}

\vfill\break

 \noindent
{\bf 1. Introduction.~~ }
\vskip 0.2cm

Starting from the seminal work of Aharonov, Bergman and Lebowitz (ABL)
(1964), Aharonov, myself and others are developing a time-symmetrized
formalism of quantum theory (TSQT). Recently a particular question
related to this formalism, namely the validity of the counterfactual
application of the ABL rule, became a subject of a significant
controversy culminating in the paper by Kastner appearing in this
issue.  According to the critics some of the recent results obtained
in the framework of the TSQT are based on the counterfactual
interpretation of the ABL rule which, in general, is inconsistent. In
recent preprints (Vaidman, 1996a, 1997) I defended the TSQT and,
moreover, introduced time-symmetrized counterfactuals for quantum
theory for which the ABL rule is valid.  Kastner critically analyzes
my preprints and claims that my defense is not sound. In this paper I
refute Kastner's arguments.  For self-consistency I include the relevant
sections of the preprints.

Lewis (1986, 34), who is probably the main authority on
counterfactuals, writes: ``Counterfactuals are infected with
vagueness, as everybody agrees.''  I do not completely agree. I believe that quantum
counterfactuals can be defined unambiguously (as I am doing in Section
3). However, it seems that the core of the current controversy is
indeed the ambiguity of the concept of counterfactuals.  Kastner
distinguishes between two apparently counterfactual readings of the
ABL rule: the first one she considers as ``non-counterfactual'' and
the second is a ``{\em bona fide} counterfactual'' (p.3(??)). She
claims that the second reading is what is frequently applied in the
TSQT.  I will argue below that it is the first reading which is
correct and which has been applied in the framework of the TSQT.  The
question whether the first reading is ``counterfactual'' remains a
semantic issue.  Formally, it is. Moreover, I will argue below that
the second Kastner's reading is inconsistent and that all other
proposed counterfactual readings of the ABL rule are either
inconsistent or not time-symmetric. Thus, I find it appropriate to use
the term ``counterfactual'' for my (Kastner's first) reading.  However, if philosophers find
it important to spell out the differences between these
``non-counterfactual'' counterfactuals and ``{\em bona fide}
counterfactuals'' applied in general analyses, I hope that the current
discussion will help in this task.

The plan of this paper is as follows.  In Section 2 I briefly define
the time-symmetrized formalism.\footnote{Sections 2-5 are from Vaidman
  (1997) and Section 6 from Vaidman (1996a). These are the preprints
  which Kastner criticizes. The full text of the preprints is
  available electronically.}  In section 3 I analyze the
concept of counterfactuals in quantum theory and introduce the
time-symmetrized counterfactuals. In Section 4 I discuss {\em elements
  of reality} which are examples of quantum counterfactuals. Section 5
is devoted to the analysis of the inconsistency proof of Sharp and
Shanks (S\&S) (1993) and its variations.  Section 6 presents more
detailed analysis of possible counterfactual interpretations of the
ABL rule.  In Section 7 I analyze Kastner's readings of the ABL rule,
in Section 8 her analysis of the S\&S proof, and in Section 9 her
criticism of my definition of time-symmetrized counterfactuals.
Section 10 summarizes the arguments of the paper.

\break

{\bf  2. Time-Symmetrized Formalism.~~}
\vskip 0.2  cm

In standard quantum theory a complete description of a system at a
given time is given by a quantum state $|\Psi \rangle$. It yields the
probabilities for all outcomes $a_i$ of a measurement at that time of any observable 
$A$ according to the equation
\begin{equation}
  \label{prob1}
  {\rm Prob}(a_i) = |\langle \Psi | {\bf P}_{A=a_i} | \Psi \rangle | ,
\end{equation}
where ${\bf P}_{A=a_i}$ is the projection operator on the subspace defined
by $A= a_i$. Although it is not manifestly apparent,  Eq. 1 is intrinsically asymmetric in time: the state
$|\Psi\rangle$ is determined by some measurements in the past and it
evolves toward the future.  The time evolution between the
measurements, however, is considered time symmetric since it is
governed by the Schr\"odinger equation for which each forward evolving
solution has its counterpart (its complex conjugate with some other
well understood simple changes) evolving backward in time. The
asymmetry in time of the standard quantum formalism is manifested in the
absence of  the quantum state evolving backward in time from 
future measurements (relative to the time in question).

Time-symmetrized quantum theory completely describes a system at a given time by
a two-state vector $\langle \Psi_2| |\Psi_1 \rangle$. It yields the
(conditional) probabilities for all outcomes $a_i$ of a measurement of
any observable $A$ at that time according to the generalization of the ABL formula
(Aharonov and Vaidman, 1991):
\begin{equation}
  \label{ABL}
 {\rm Prob}(a_i) = {{|\langle \Psi_2 | {\bf P}_{A=a_i} | \Psi_1 \rangle |^2}
\over{\sum_j|\langle \Psi_2 | {\bf P}_{A=a_j} | \Psi_1 \rangle |^2}} .
\end{equation}

 The time symmetry means that $\langle \Psi_2|$ and $|\Psi_1
\rangle$ enter the equations, and thus govern the observable results,
on equal footings.
Moreover, the time symmetry  means that, in regard to time symmetric
measurements, a system described by the
two-state vector $\langle \Psi_2| |\Psi_1 \rangle$ is identical to a system described by the
two-state vector $\langle \Psi_1| |\Psi_2 \rangle$.
I  analyze the time symmetry of the process of measurement in
section 6 of (Vaidman, 1997), here I only point out that ideal measurements are time
symmetric. Indeed, the  symmetry under the interchange of $\langle
\Psi_2|$ and  $|\Psi_1 \rangle$ is explicit in Eq. 2 which refers to
ideal measurements.

Another basic concept of time-symmetrized two-state vector formalism is
{\em weak value}. An  (almost) standard measurement
procedure  for measuring observable $A$ with weakened coupling (which we call {\em weak
  measurement}, Aharonov and Vaidman 1990) yields  the {\em weak
  value} of $A$:
\begin{equation}
A_w \equiv { \langle{\Psi_2} \vert A \vert\Psi_1\rangle 
\over \langle{\Psi_2}\vert{\Psi_1}\rangle } .
\label{wv}
\end{equation}
Here again, $\langle \Psi_2|$ and $|\Psi_1
\rangle$ enter the equations
on equal footings. However, when we interchange $\langle \Psi_2|$ and $|\Psi_1 \rangle$, the weak
value changes to its complex conjugate. Thus, in this situation, as for
the Schr\"odinger equation, time reversal is accompanied by complex
conjugation.

In order to explain how to obtain a quantum system described at a
given time $t$ by a two-state vector $\langle \Psi_2| |\Psi_1 \rangle$
we shall assume for simplicity that the free Hamiltonian of the system
is zero. In this case, it is enough to prepare the system at time
$t_1$ prior to  time $t$ in the state $|\Psi_1\rangle$, and to ensure
no disturbance between $t_1$ and $t$ as well as between $t$ and $t_2$,
and to find the system at $t_2$ in the state $|\Psi_2\rangle$. It is
crucial that $t_1 < t < t_2$, but the relation between these times and
``now'' is not fixed. The times $t_1, t, t_2$ might all be in the past, or we
can discuss future measurements and then they are all in the
future; we just have to agree to discard all cases when the
measurements at  time $t_2$ does not yield the outcome
corresponding to the state $|\Psi_2\rangle$.

Note the asymmetry between the measurement at $t_1$ and the
measurement at $t_2$. Given an ensemble of quantum systems, it is
always possible to prepare all of them in a particular state
$|\Psi_1\rangle$, but we cannot ensure finding the system in a
particular state $|\Psi_2\rangle$.  Indeed, if the pre-selection
measurement yielded a result different from projection on
$|\Psi_1\rangle$ we can always change the state to $|\Psi_1\rangle$,
but if the measurement at $t_2$ did not show $|\Psi_2\rangle$, our
only choice is to discard such a system from the ensemble.
 This  asymmetry,
however, is not relevant to the problem we consider here.  We study
the symmetry relative to the measurements at time $t$ for a {\em given} pre-
and post-selected system, and we do not investigate the time-symmetry
of obtaining such a system. The only important detail is that the
interaction  at time $t$ has to be time symmetric.
 See more discussion  in Section 6 of (Vaidman, 1997).

\vskip 1.cm \noindent
{\bf 3. Counterfactuals.~~}
\vskip 0.2cm

 A general form of a counterfactual statement
is
\begin{quotation}
 { \bf (i)}
{\em If it were that $\cal A$, then it would be that $\cal B$.}
\end{quotation}
There are  many philosophical discussions on the concept of
counterfactuals and especially on  time's arrow in counterfactuals.
Many of the discussions, e.g. Lewis (1986), Bennett (1984), are
related to $\cal A$: How come $\cal A$ if in the actual world $\cal A$ is
not true? Do we need a miracle (a violation of the fundamental law
of nature) for $\cal A$?  Does $\cal A$ come
by itself, or it is accompanied by other changes? However, these
questions are not relevant to the problem of counterfactuals in
quantum theory.  The questions about $\cal A$ are not relevant because
$\cal A$ depends solely on an external system which is not under
discussion by the definition of the problem.  Indeed, in quantum theory
the counterfactuals have a very specific form:\footnote{This
  definition of counterfactuals in quantum theory is broad enough for
  discussing issues relevant to this paper. However, in some cases the
  term ``counterfactuals'' has been used differently. For example, in
  Penrose (1994, p.240) ``{\em counterfactuals} are things that might
  have happened, although they did not in fact happen.''}
\begin{quotation}

 $\cal A$ =  a measurement  $\cal M$   is performed

 $\cal B$ = the outcome of $\cal M$  has  property $\cal P$
\end{quotation}
The measurement $\cal M$  might consist of
measurements of several observables  performed together. The
property $\cal P$ might be a certain
relation between the results of measurements of these observables or a
probability for a certain relation or for a certain outcome.

It is assumed that the experimenter can make any decision about which
measurement to perform and the question how he makes this decision is
not considered. It is assumed that the experimenter and his measuring
devices are not correlated in any way with the state of the system
prior to the measurement. Thus, in the world of the quantum system no
miracles are needed and no changes relative to the actual world have to be
made for different $\cal A$'s.\footnote 
  {Indeterminism of standard
  quantum theory allows us to discuss even the worlds which include the
  experimenter without invoking miracles. Consider an experimenter who
  chooses between different measurements according to a random outcome
  of another quantum experiment.}

Although one can define counterfactuals of this form in the framework
of classical theory, they are of no interest because they are
equivalent to some ``factual'' statements. In classical physics any
observable has always a definite value and a measurement of the observable
yields this value. Therefore, we can make a one to one correspondence
between ``the outcome of a measurement of an observable $C$ is $c_i$''
and ``the value of $C$ is $c_i$''. The latter is independent of
whether the measurement of $C$ has been performed or not and,
therefore, statements which are formally counterfactual about results
of possible measurements can be replaced by ``factual''
(unconditional) statements about values of corresponding observables.
In contrast, in standard quantum theory, observables, in general, do not
have definite values and therefore we cannot always reduce the above
counterfactual statements to ``factual'' statements.

Most of the discussions of counterfactuals in quantum theory are in
the context of EPR-Bell type experiments. Some of the examples are
Skyrms (1982), Peres (1993), Mermin (1989) (which, however, does not
use the word counterfactual), Ghirardi and Grassi (1994) and Bedford
and Stapp (1995) who even present an analysis of a Bell-type argument
in the formal language of the Lewis (1973) theory of counterfactuals.
The common situation is that a composite system is described at a
certain time by some entangled state and then an array of incompatible
measurements on this system at a later time is considered. Various
conclusions are derived from statements about the results of these
measurements. Since these measurements are incompatible they cannot be
all performed together, so it must be that at least some of them were
not actually performed.  This is why they are called counterfactual
statements.

These counterfactuals are explicitly asymmetric in time. The
asymmetry is neither in $\cal A$ nor in $\cal B$; both are about a
single time  $t$.
 The asymmetry is in the description of the actual world.
 The {\it past} and not the {\it
  future}  (relative to $t$) of a system is given. 

This, however, is not the only asymmetry of the counterfactuals in
quantum theory as they are usually considered. A different asymmetry
(although it looks very similar) is in what we assume to be ``fixed'',
i.e.,  which properties of the actual world we assume to be true in
possible counterfactual worlds.  The {\it past} and not the {\it
  future} of the system is fixed.

It seems that while the first asymmetry can be easily removed, the
second asymmetry is unavoidable. According to standard quantum theory
a system is described by its quantum state. In the actual world, in
which a certain measurement has been performed at time $t$ (or no
measurement has been performed at $t$) the system is described by a
certain state before $t$, and by some state after time $t$. In the
counterfactual world in which a different measurement was performed at
time $t$, the state before $t$ is, of course, the same, but the state
after time $t$ is invariably different (if the observables measured in
actual and counterfactual worlds have different eigenstates).
Therefore, we cannot hold fixed the quantum state of the system in the
future.\footnote {Note that none of these asymmetries exists in the
  classical case because when a complete description of a classical
  system is given at one time, it yields and fixes the complete
  description at all times and (ideal) measurements at time $t$ do not
  change the state of a classical system.}

The argument above shows that for  constructing  time
symmetric counterfactuals we have to give up the description of a
quantum system by its quantum state. Fortunately we can do that without loosing
anything except the change due to the measurement at time $t$ which
caused the difficulty. A quantum state at a given time is completely
defined by the results of a complete set of measurements performed
prior to this time.  Therefore, we can take the set of all results
performed on a quantum system as a description of the world of the
system instead of describing the system by its quantum state. (This
proposal will also help to avoid ambiguity and some controversies
related to the description of a single quantum system by its quantum
state.) Thus, I propose the following  definition of
counterfactuals in the framework of quantum theory:

\begin{quotation}
{ \bf (ii)}
{\em  If  a measurement ${\cal M}$  were performed at
  time $t$, then it would have property ${\cal P}$, provided  that the results of all measurements
performed on the system at all times except the time $t$ are fixed.}
\end{quotation}
 
For time asymmetric situations in which only the results of
measurements performed before $t$ are given (and thus only these
results are fixed) this definition of counterfactuals is equivalent to
the counterfactuals as they usually have been used. However, when the
results of measurements performed on the system both before and after
the time $t$ are given, definition (ii) yields novel time-symmetrized
counterfactuals. In particular, for the ABL case, in which {\em
  complete} measurements are performed on the system at $t_1$ and
$t_2$, $t_1 < t <t_2$, we obtain
\begin{quotation}
{ \bf (iii)}
{\em  If  a measurement of an observable $C$ were performed at
  time $t$, then  the probability for
$C=c_i$ would equal $p_i$, provided  that the results of  measurements
performed on the system at times $t_1$ and $t_2$  are fixed.}
\end{quotation}
The ABL formula (2) yields correct probabilities for counterfactuals
defined as in (iii), i.e., in the experiment in which $C$ is measured at
time $t$  on the systems from a pre- and
post-selected ensemble defined by fixed outcomes of the measurements at
$t_1$ and $t_2$ (all such   systems and only such systems are
considered)
the frequency of an outcome $c_i$ is  $p_i$.

 For  the ABL situation  one
can also define a time
{\em asymmetric} counterfactual:
\begin{quotation}
{\bf (iv)} Given the results of measurements at $t_1$ and $t_2$, $t_1 < t
<t_2$ (in the actual world),
if  a measurement of an  observable $C$ were performed at
  time $t$, then  the probability for
$C=c_i$ would equal $p_i$, provided  that the results of all measurements
performed on the system at all times before  time $t$ are fixed.
\end{quotation}
In the framework of standard quantum theory the information about the result
of measurement at $t_2$ is irrelevant: the probability for
$C=c_i$ does not depend on this result. Thus, it is obvious that the
ABL formula (2), which includes the result of the measurement at time
$t_2$ explicitly,  does not yield counterfactual probabilities according
to definition (iv).

 One might modify definition (iv) in the framework of
some ``hidden variable'' theory with a natural additional requirement of
fixing the hidden variables of the system in the past. The properties of
such counterfactuals will depend crucially on the details of the hidden
variable theory (see the discussion of Aharonov and Albert (1987) in the
framework of Bohm's theory), but  the ABL formula (2) is not valid
for any such modification. 
In order to show this consider a spin-${1\over 2}$ particle which  was found  at
$t_1$ and  at  $t_2$ in the same state
$|{\uparrow}_z\rangle$
 (and no measurement has been performed at $t$). 
 We ask what is the (counterfactual) probability for finding
spin ``up'' in the direction $\hat \xi$ which makes an angle $\theta$
with the direction $\hat z$, at the intermediate time $t$. In this case,
 hidden variables, even if  they exist, cannot change that probability
because any   particle found at $t_1$ in the state $|{\uparrow}_z\rangle$, irrespectively of
its hidden variable, 
yields the outcome ``up'' in the 
measurement at  $t_2$. Therefore, the statistical predictions
about the intermediate measurement at time $t$ must be the same as for
the pre-selected only ensemble (these are {\em identical} ensembles in
this case), i.e.
\begin{equation}
  \label{abl-qm}
  {\rm Prob}({\uparrow}_\xi) =|\langle {\uparrow}_\xi |
{\uparrow}_z\rangle|^2 = \cos^2(\theta/2).
\end{equation}
The ABL formula, however, yields: 
\begin{equation}
  \label{abl-xiz}
  {\rm Prob}({\uparrow}_\xi) = {{|\langle {\uparrow}_z |
      {\bf P}_{{\uparrow}_\xi} | {\uparrow}_z \rangle |^2}\over{|\langle
      {\uparrow}_z | {\bf P}_{{\uparrow}_\xi} | {\uparrow}_z \rangle |^2
    +|\langle {\uparrow}_z | {\bf P}_{\downarrow_\xi} | {\uparrow}_z \rangle
  |^2}}= {{ \cos^4(\theta/2)}\over{ \cos^4(\theta/2) +  \sin^4(\theta/2)}} .
\end{equation}
The fact that the ABL formula (2) does not hold for counterfactuals
defined in (iv) or its modifications is not surprising. Definition
(iv) is explicitly asymmetric in time. The ABL formula, however, is
time symmetric and therefore it can hold only for time-symmetrized
counterfactuals.

A recent study of time's arrow and counterfactuals in the framework of
quantum theory by Price (1996) seems to support my definition (ii).
Let me quote from his section ``Counterfactuals: What should we
fix?'':
\begin{quotation}
Hold fixed the past, and the same difficulties arise all over
again. Hold fixed merely what is accessible, on the other hand, and it
will be difficult to see why this course was not chosen from the
beginning. (1996, 179)
\end{quotation}
This quotation looks very much like my proposal. Indeed, I find
many arguments  in his book pointing in the same direction. However, 
in fact,  this quotation represents a time asymmetry: according to Price
``merely what is accessible''  is ``an accessible
past''. But this is not the time asymmetry of the physical theory;   Price writes: ``no physical asymmetry is
required to explain it.''  Although the book includes an extensive
analysis of a photon passing through two polarizers -- the classic
setup for the ABL case, I found no
explicit discussion of a possible measurement in between, the problem we discuss here.\footnote{Price briefly and critically mentions the ABL paper. He
  writes (1996, 208): ``What they [ABL] fail   to note, however, is
  that  their argument  does nothing to address the problem for those
 who disagree with Einstein -- those  who think that the state
 function is a complete description, so that the change that takes
 place on measurements is a real change in the world, rather than
 merely change in our knowledge of the world.'' This seems to me an
 unfair criticism:  ABL clearly state that in the situations they
 consider ``the complete description'' is given by  {\it two} wave
 functions (see more in Aharonov and Vaidman 1991). Moreover, it seems
 to me that the  development of this time-symmetrized quantum formalism
 is not too far from the spirit of the ``advanced action'' -- the Price
 vision of the solution of the time's arrow problem.}

\vskip 1.cm \noindent
{\bf 4. Elements of Reality.~~}

\vskip 0.2cm

  Important counterfactuals in quantum
theory are  ``elements of reality''. For comparison, I'll start with a
definition of time asymmetric element of reality: 
 \begin{quotation}
 {\bf (v)~}   If we can {\it predict} with certainty that the result of
   measuring at time $t$ of an observable $A$ is $a$, then, at time
   $t$, there exists an element of reality $A=a$.
  \end{quotation}
  This is, essentially, a quotation from Redhead (1987), who, however
  considered it as a sufficient condition and not as a definition.
  Redhead was inspired by the criteria for elements of reality of
  Einstein, Podolsky and Rosen (EPR). In spite of similarity in its
  form, the EPR criteria, taken as a definition, is very different:
  ``If, without in any way disturbing the system, we can predict with
  certainty the value of a physical quantity...''  The crucial
  difference is that ``predict'' in the EPR definition means to find
  out using certain (non-disturbing) measurements, while in my
  definition ``predict'' means to deduce using existing information.
  Thus, for two spin-${1\over 2}$ particles in a singlet state, the
  value of a spin component of a single particle in any direction is
  an element of reality in the EPR sense (it can be found out by
  measuring another particle) and there is no element of reality for
  a spin component value in any direction according to my definition
  (in the EPR state, the probability to find spin ``up'' in any direction
  is ${1\over 2}$).

 Definition (v) of elements of reality is
  asymmetric in time because of the word ``predict''. I have proposed
  a modification of this definition applicable for time symmetric
  elements of reality (Vaidman 1993):
 \begin{quotation}
 {\bf (vi)~}   If we can {\em infer} with certainty that the result of
   measuring at time $t$ of an observable $A$ is $a$, then, at time
   $t$, there exists an element of reality $A=a$.
\end{quotation}
The word ``infer'' is neutral relative to past and future. The
inference about results at time $t$ is based on the results of
measurements on the system performed both before and after time $t$.
Note, that in some situations we can ``infer'' more facts than can be
obtained by ``prediction'' based on the results in the past and
``retrodiction'' based on the results in the future (relative to $t$)
together.

The difference between definitions of ``elements of reality'', (v) and
(vi), and definitions of counterfactuals in quantum theory (iv) and
(iii) is that the property ${\cal P}$ in (v) and (vi) is constrained
to ``the result of measuring at time $t$ of an observable $A$ is
$a$''.  In fact, time asymmetric ``elements of reality'' (v), defined
as ``predictions'', do not represent ``interesting'' counterfactuals.
There is no nontrivial set of such counterfactual statements, i.e.,
set of statements which cannot be tested all on a single system.
Indeed, all observables the measurement of which yield definite
outcomes for a pre-selected system can be tested together.  One way to
extend the definition of time asymmetric elements of reality in order
to get nontrivial counterfactuals is to consider ``multiple-time
measurements'' (instead of measurements at time $t$ only). Another
extension, which corresponds to numerous analyses in the literature,
is to go beyond statements about observables which have definite
values:
 \begin{quotation}
  {\bf (vii)~}    If we can {\it predict} with certainty a certain relation
   between  the results $a_j$ of
   measuring at time $t$ a set of observables $A_j$, then, at time
   $t$, there exists a ``generalized element of reality'' which is this
   relation between $a_j$'s.
  \end{quotation}

A simple example of this kind is a system of  two spin-${1\over 2}$
particles prepared, at
$t_1$, in a singlet
state 
\begin{equation}
|\Psi_1\rangle = {1\over {\sqrt 2}}(  |{{\uparrow}}\rangle_1
|{\downarrow}\rangle_2 -   |{\downarrow}\rangle_1
|{{\uparrow}}\rangle_2).
\end{equation}
 We can predict with certainty that 
the results of measurements of spin components of the two particles
fulfill the following two relations:
\begin{eqnarray} 
\{\sigma_{1x}\} + \{\sigma_{2x}\} = 0 , \\
\{\sigma_{1y}\} +\{ \sigma_{2y}\} = 0 ,
\end{eqnarray} 
where $\{\sigma_{1x}\}$ signifies the result of measurement of spin $x$
component of the first particle, etc.  The relations (7) and (8)
represent a set of generalized elements of reality (vii). This is a
nontrivial set of counterfactuals because (7) and (8) 
cannot be tested together: the measurement of $\sigma_{1x}$
disturbs the measurement of $\sigma_{1y}$ as well as the measurement
of $\sigma_{2x}$ disturbs the measurement of $\sigma_{2y}$.

In contrast,   the set of elements of reality  
(vi) given by 
\begin{eqnarray} 
\{\sigma_{1x} + \sigma_{2x}\} = 0 , \\
\{\sigma_{1y} + \sigma_{2y}\} = 0 ,
\end{eqnarray} 
can be tested on a single system, see
  Aharonov et al. (1986) for description of such measurements.
Yet another set of counterfactuals, which consists of definite
statements about measurements, but which does not fall
into category (vi) because these are {\em two-time}  measurements performed at two
different time moments $t_1$ and $t_2$, {\em cannot} be tested on a
single system:
\begin{eqnarray} 
\{\sigma_{1x}(t_1) + \sigma_{2x}(t_2)\}  = 0 , \\
\{\sigma_{1y}(t_2) + \sigma_{2y}(t_1)\} = 0 .
\end{eqnarray} 
Note a  situation which involves only a single free
spin-${1\over 2}$ particle. The particle is  prepared, before $t_1$, $t_1 < t_2 < t_3$, in the state
$|{{\uparrow}}_y\rangle$. Then, a nontrivial set of counterfactuals is:
\begin{eqnarray} 
\{\sigma_{x}(t_1) - \sigma_{x}(t_3)\} = 0 , \\
\{\sigma_{y}(t_2)\}  = 1 .
\end{eqnarray} 
 In this example, however, statement (13) has somewhat different
character because  
it  depends not   on the results of measurements performed on the
particle before or after the period of time $(t_1, t_2)$, but on the
fact that the system was not disturbed during this period of time.

\break 
{\bf 5. Inconsistency proofs.~~}
\vskip 0.2cm

The key point of the criticism of the time-symmetrized quantum theory (Sharp and Shanks 1993; Cohen 1995; Miller 1996)
is the conflict between counterfactual interpretations of the ABL rule
and predictions of quantum theory. I shall argue here that the inconsistency proofs
 are unfounded and therefore the criticism
essentially breaks  apart.

The  structure of all these inconsistency proofs is a follows.
  Three consecutive measurements are considered. The
first is the preparation of the state $|\Psi_1\rangle$ at  time
$t_1$.  The probabilities for the results $a_i$ of the second
measurement at  time $t$ are considered. The final measurement
at  time $t_2$ is introduced in order to allow the analysis 
using the ABL formula.  Sharp and Shanks consider three consecutive
spin component measurements of a spin-${1\over 2}$ particle in different
directions. Cohen analyzes a particular single-particle
interference experiment. It is a variation  of
the Mach-Zehnder interferometer with two detectors for the final
measurement and the possibility of placing  a third detector for the
intermediate measurement. Finally, Miller repeated the argument for a
 system of tandem Mach-Zehnder interferometers.
  In all these cases the ``pre-selection only''
situation is considered.
It is unnatural to apply the time-symmetrized formalism for such
cases. However, it must be possible.  Thus, I need not show that the
time-symmetrized formalism has an advantage over the standard
formalism for describing these situations, but   only that it is 
consistent  with the predictions of the standard quantum theory. 

In the standard approach to quantum theory the
probability for the result of a measurement of $A$ at  time $t$
is given by Eq. 1. The claim of all the proofs is that the counterfactual
interpretation of the ABL rule yields a different result.  In all cases
the final measurement at time $t_2$ has two possible outcomes
which we signify as ``$1_f$'' and ``$2_f$''.
The suggested application of the ABL rule is as follows. The
probability for the result $a_i$ is:
\begin{equation}
  \label{p1}
  {\rm Prob}(A=a_i)~=~  {\rm Prob}(1_f) ~ {\rm Prob}(A=a_i |1_f) +
  {\rm Prob}(2_f) ~{\rm Prob}(A=a_i |2_f),
\end{equation}
where $ {\rm Prob}(A=a_i |1_f)$ and $ {\rm Prob}(A=a_i |2_f)$ are the 
conditional probabilities given by the ABL formula, Eq. 2, and $ {\rm
  Prob}(1_f)$ and $ {\rm Prob}(2_f)$ are the probabilities for the
results of the final measurement.
In the proofs, the authors show that Eq. \ref{p1} is not valid and conclude
that  the ABL formula is not applicable for this example and 
therefore that it is not applicable in general.

I will argue that the error in calculating equality (\ref{p1}) is not in the conditional
probabilities given by the ABL formula, but   in the calculation of the probabilities $ {\rm
  Prob}(1_f)$ and $ {\rm Prob}(2_f)$ of the final measurement. In all three
cases it was calculated on the assumption that {\rm no} measurement
took place at  time $t$. Clearly, one cannot make this assumption
here since then the discussion about the probability of the result of
the measurement at  time $t$ is meaningless. Unperformed
measurements have no results (Peres, 1978). Thus, there is no  surprise that the
value for the probability ${\rm Prob}(A=a_i)$ obtained in this way
comes out different from the value predicted by the quantum theory.

Straightforward calculations show that if one uses the formula (\ref{p1})
with the probabilities $ {\rm Prob}(1_f)$ and $ {\rm Prob}(2_f)$
calculated on the condition that the intermediate measurement has been
performed, then the outcome is the same as predicted by the standard
formalism of quantum theory.  Consider, for example, the experiment
suggested by Sharp and Shanks, consecutive spin measurements with
the three directions in the same plane and the relative angles
$\theta_{ab}$ and $\theta_{bc}$. The probability for the final result
``up'' is

\begin{equation}
  \label{p1f}
  {\rm Prob}(1_f) =
\cos^2(\theta_{ab}/2)\cos^2(\theta_{bc}/2)+
\sin^2(\theta_{ab}/2)\sin^2(\theta_{bc}/2),
\end{equation}
  and the probability for
the final result ``down'' is
\begin{equation}
  \label{p12}
 {\rm Prob}(2_f)
=\cos^2(\theta_{ab}/2)\sin^2(\theta_{bc}/2)+
\sin^2(\theta_{ab}/2)\cos^2(\theta_{bc}/2). 
\end{equation}
 The ABL formula yields
\begin{equation}
  \label{p1abl}
 {\rm Prob}(up |1_f)={{\cos^2(\theta_{ab}/2)\cos^2(\theta_{bc}/2)}
  \over {\cos^2(\theta_{ab}/2)\cos^2(\theta_{bc}/2)+
    \sin^2(\theta_{ab}/2)\sin^2(\theta_{bc}/2)}}
\end{equation}
 and 
\begin{equation}
  \label{p2abl}
 {\rm
  Prob}(up |2_f)={{\cos^2(\theta_{ab}/2)\sin^2(\theta_{bc}/2)}
  \over {\cos^2(\theta_{ab}/2)\sin^2(\theta_{bc}/2)+
    \sin^2(\theta_{ab}/2)\cos^2(\theta_{bc}/2)}}.
\end{equation}
 Substituting all
these equations into Eq. \ref{p1} we obtain
\begin{equation}
  \label{paiabl}
 {\rm
  Prob}(up)=\cos^2(\theta_{ab}/2).
\end{equation}
This result coincide with the prediction of the standard quantum
theory. It is a straightforward exercise to show in the same way that
no inconsistency arises also in the examples of Cohen\footnote{In Cohen's
  example the measurement at time $t_2$ is not a complete measurement
  and therefore the ABL formula (2) is not applicable for this
  case. The analysis requires a generalization of the ABL formula
  given in Vaidman (1998a).} and Miller.

I have shown that one can apply the time-symmetrized formalism, including
the ABL formula, for analyzing the examples which allegedly lead to
contradictions in the inconsistency proofs. In my analysis there was
nothing ``counterfactual''. The proofs, however, claimed to show that
a ``counterfactual interpretation'' of the ABL rule leads to
contradiction.
 What I have shown is that the examples
presented in the proofs do not correspond to counterfactual situations
and this is why they cannot be analyzed in a counterfactual way. The
contradictions in the proofs arise from a logical error in taking
together the statement ``no measurement has been performed at $t$''
and a statement about probability of a result of this measurement
which requires ``the measurement has been performed at $t$''.
Let me demonstrate how  similar erroneous ``counterfactual''
reasoning  can  lead to a
contradiction in quantum theory even in cases when the ABL rule is not
involved.  Consider two consecutive measurements of $\sigma_x$
performed on a spin-${1\over 2}$ particle prepared in a state
$|{\uparrow}_z\rangle$. Let us ask (using the language of Sharp and
Shanks) what is the probability that these measurements {\it would
  have had} the results $\sigma_x(t_1) =\sigma_x(t_2) =1$ given that
no such measurements in fact took place. Each spin measurement, if
performed separately, has probability ${1\over 2}$ for the result $\sigma_x
=1$. According to standard quantum theory the fact that  in the actual world the
measurement at $t_1$ has been performed and $\sigma_x(t_1)=1$ has been
obtained  does not ensure  that in a counterfactual world in which
$\sigma_x$ was not measured at $t_1$, but at a later time $t_2$, the
outcome has to be 
$\sigma_x(t_2)=1$, rather we still have probability $1\over 2$ for this
result. Thus, counterfactual reasoning leads us to the erroneous result
that the probability for $\sigma_x(t_1) =\sigma_x(t_2) =1$ is ${1\over
  2}\times {1\over 2} ={1\over 4}$.

\vskip 1.cm \noindent {\bf 6.  Counterfactual Interpretations of the
  ABL Probability Rule.}\footnote{This section is, in fact, a preliminary
  version of Section 3. I  bring it here because it discusses
  several points in more details and mostly because Kastner refers
  explicitly to
  the text of this section.}
 
\vskip 0.2cm

 In this section I shall consider three ways
to interpret the ``counterfactual interpretation''. The first
interpretation I cannot comprehend, but I have to discuss it since it
was proposed and used in the criticism of the time-symmetrized quantum
theory. I believe that I understand the meaning of the second
interpretation, but I shall argue that it is  not appropriate
for the problem which is discussed here. The last interpretation is
the one I want to adopt and I shall present several arguments in its
favor.

\vskip .2cm
\noindent
{\bf Interpretation (a)~~}
{\it  Counterfactual probability as the probability of the result of a
measurement which has not been performed.}

Let me quote Sharp and Shanks: 
\begin{quotation}
  ...for, conditionalizing upon specified results of measurements of
  $M_I$ and $M_F$, there is no reason to assign the same values to the
  following probabilities: the probability that an intervening
  measurement of $M$ had the result $m^j$ given that such a
  measurement in fact took place, and the probability that intervening
  measurement {\em would have had} the result $m^j$ given that no such
  intervening measurement of $M$ in fact took place. In other worlds
  there is no reason to identify ${\rm Prob}(M= m^j|{\rm \bf
    E}_M[\psi^i_I, \psi^k_F])$ and ${\rm Prob}(M= m^j|{\rm \bf
    E}[\psi^i_I, \psi^k_F])$.(1993, 491)
\end{quotation}
I can not comprehend the  meaning of the probability for the result
$M=m^j$ given that the measurement $M$ has not take place. As far as I
can see $ {\rm Prob}(M= m^j|{\rm \bf  E}[\psi^i_I, \psi^k_F])$ has no
physical meaning. Sharp and Shanks  continue:
\begin{quotation}
 (For a classical illustration, consider a drug which, if injected to
 facilitate a medical test at $t$, has an effect, starting shortly after
 the test and persisting past $t_F$, on the the value of the tested
 variable. Suppose that it is unknown whether a test was conducted at
 $t$, but that a value for the tested variable is obtained at $t_F$. Using
 the value  at $t_F$, we would estimate differently the value prior to
 $t$ depending on whether we assume that a test did or did not take
 place at $t$.)
\end{quotation}
This might explain what they have in mind, but the argument does not
hold since in many situations there is no  quantum
mechanical counterpart to the classical case of ``the value [of a tested
variable] prior to  $t$'' . In standard quantum
theory {\em unperformed experiments have no results}, see Peres (1978).

Cohen and Hiley  partially acknowledge the problem admitting
that at least in the framework of the orthodox interpretation this is
a meaningless concept:
\begin{quotation}
  In other words we cannot necessarily assume that the ABL rule will
  yield the correct probabilities for what the results of the
  intermediate measurements {\em would} have been, {\em if} they had
  been carried out, in cases where these measurements have not {\em
    actually} been carried out. In fact, this sort of counterfactual
  retrodiction has no meaning in the orthodox (i.e., Bohrian)
  interpretation of quantum mechanics, although it can legitimately be
  discussed within the standard interpretation and
  within some other interpretations of quantum mechanics (see, for
  example, Bohm and Hiley[1993]).(1996, 3)
\end{quotation}
I fail to understand the interpretation (a) in any framework.
Maybe, if we restrict ourselves to the cases in which the system at
the intermediate time is in an eigenstate of the variable which we
intended to measure, (but we had not), we can associate the
probability 1 with such unperformed measurements. This is close to
the idea of Cohen (1995) to consider counterfactuals in the restricted
cases corresponding to consistent histories introduced by Griffith
(1984). But, as far as I can see, interesting situations do not
correspond to consistent histories, and therefore no novel (relative
to classical theory) features of quantum theory can be seen in this
way.  It is possible that what Cohen and Hiley (1996) have in
mind is the interpretation (b) which I shall discuss next.

\vskip .2cm
\noindent
{\bf Interpretation (b)}

{\it Counterfactual probability as the probability of the result of a
measurement would it have been performed based on the information
about the world in which the measurement has not been performed.}

At time $t_1$ we preselect the state $|\Psi_1\rangle$. We do not
perform any measurement at time $t$. We perform a measurement at time
$t_2$ and find the state $|\Psi_2\rangle$. We ask, what would be the
probability for the results of a measurement performed at time $t$ in
a world which is identical to the actual world at time $t_1$.

This is a meaningful concept, but I believe that it is not adequate
for discussing pre- and post-selected quantum systems because it is
explicitly asymmetric in time. The counterfactual world is identical to the
actual world at time $t_1$ and might not be identical at time $t_2$.

[This interpretation is equivalent to (iv) of Section 3, see discussion
there.]

\vskip .2cm
\noindent
{\bf Interpretation (c)~~}
{\it Counterfactual probability as the probability for the results of
  a measurement if it has been performed in the world ``closest'' to
  the actual world.}

This  
is identical in  form and  spirit to the theory of
counterfactuals of Bennett (1984), although the context of  the pre- and
post-selected quantum measurements is somewhat beyond what he
considered. This interpretation is explicitly time-symmetric. The
title, however,  does not specify  it completely and I shall explain what I
mean (in particular by the word ``closest'') now.

I have to specify the concept of ``world''. There are many parts of
the world which do not interact with the quantum system in question,
so their states are irrelevant to the result of the measurement. In our
discussion we might include all these irrelevant parts, or might not,
without changing any of the conclusions.  There are other aspects of
the world which are certainly relevant to the measurement at time $t$,
but we postulate that they should be  disregarded. Everything which is connected to
our decision to perform the measurement at time $t$ and all the
records of the result of that measurement are not considered.  Clearly,
the counterfactual world in which a certain measurement has been
performed is different from an actual world in which, let us assume, no
measurement has been performed at time $t$. The profound differences
are both in the future where certain records exist or do not exist and in
the past which must be different since one history leads to performing
the measurement at  time $t$ and another history leads to no
measurement.\footnote{If a random process chooses between the two
possibilities, then the past before this process might be identical.}
However, our decision to make the measurement is not connected to the
quantum theory which makes predictions about the result of that
measurement. We want to limit ourselves to the discussion of the
time-symmetry of the quantum system. We do  not consider here the question of the
time-symmetry of the entire  world. Therefore,  we  exclude the
external parts from our
consideration.

What constitutes a description of a quantum system itself is also a very
controversial subject.  The reality of the Schr\"odinger wave,
the existence or nonexistence of hidden variables etc. are subjects of
hot discussions. However, everybody agrees that the collection of all
results of measurements is a consistent (although maybe not complete)
description of the quantum system. Thus, I propose the following
definition:
\begin{quotation}
{\em A world ``closest'' to the actual world is  a world in which all
measurements (except the measurement at the time  $t$ if performed)
have the same outcomes as in the actual world.}
\end{quotation}
This definition overcomes the common objection according to which one
should not consider together statements about pre- and post-selected
systems regarding different measurements at time $t$ because these
systems belong to different ensembles. The difference is in their
quantum state at the time period between $t$ and $t_2$.\footnote{If one is
adopting our backward evolving quantum state, he can add that the
systems are also different due to the backward evolving state between
$t$ and $t_1$.} Formally, the problem is solved by considering only
results of measurements and not the quantum state.
The justification of this step follows from the rules of the game: it is postulated that the quantum system is not disturbed
during the periods of time $(t_1, t)$ and $(t, t_2)$. Therefore, it is
postulated that no measurement on the system is performed during these
periods of time. Since unperformed measurements  have no results, the
difference between the ensembles has no physical meaning in the
discussed problem.

From the alternatives I have presented here, only interpretation (c) is
time-symmetric. This is the reason why I believe that it is the only
reasonable candidate for analyzing the (time-symmetric) problem of
measurements performed between two other measurements. [Interpretation
(c) is equivalent to interpretation (iii) of Section 3.]

\vskip 1.cm \noindent {\bf 7. Kastner's readings of the ABL rule} 
\vskip 0.2cm 
Kastner  puts in quotation marks two possible readings of the ABL rule,
``non-counterfactual'' and ``{\em bona-fide} counterfactual'' (1999,
p. 3[??]). 
As far as I can see the two quotations are not different: the first is
a clarification of the second.
In the papers of the TSQT the statements frequently appear in the
compact form of the second quotation and the first quotation is the
correct explanation  of their meaning.
The explicit notation of the ABL
rule of Kastner's Eq. $1'$ is the correct characterization of both
quotations.

Kastner, however, reads the second quotation differently.  She claims
that there is a ``quantifier ambiguity'' in the ABL formula (her Eq.
1). She proposes to add a parameter, $C$, indicating the variable
which was actually measured at time $t$. It is not clear what Kastner
means by ``actually'', and this is crucial for her arguments (see
especially her footnote 3).  

The first possible reading of Kastner is
that $C$ is related to the counterfactual world, for which the formula
should yield probabilities for $x_j$. This reading is equivalent to
interpretation (a) of Section 6, and as I showed there it is
meaningless. In the framework of quantum theory observables usually do
not possess values.  There is no meaning for ``probability of a
value'', only for ``probability of an outcome of a measurement''. If
it is postulated that $X$ is not measured (since another variable,
$C$, is measured instead), then it is meaningless to ask what is the
probability for $x_j$. In other words, the question is what parameters
are kept fixed when the counterfactual world is considered. Kastner's
notation, $P(x_j|a,b;C)$, suggests that $a,b$, and $C$ are kept fixed
in the counterfactual world, but then there is no meaning for
probability of $x_j$. 

The second possible reading of Kastner (and I understand from the
correspondence with her that this is the correct reading) is that $C$
relates to the actual world and the formula is related to a
counterfactual world, in which another variable, $X$, is measured. In
this case it is not clear what is kept fixed in the counterfactual
world. If $a$ and $b$ are kept fixed, then how $C$ can be relevant?
The question is about the counterfactual world which is specified
completely by $a$ and $b$, so the information about what has happened
in another (the actual) world is irrelevant.  Finally, it might be
that Kastner assumes some hidden variables which kept fixed, i.e.,
identical for actual and counterfactual worlds. Then $C$ is relevant
because it characterizes the hidden variables: they are such that,
given the intermediate measurement $C$, the outcome $b$ is obtained.
This is a modification of interpretation (iv) of Section 3
(interpretation (b) of Section 6),  in the
latter  $C$
is the identity $I$. While it is not immediately obvious that the ABL
formula fails for interpretation (iv) (it does fail as it is proved in
Section 3), it is obvious that the ABL formula cannot be true for this
particular reading of Kastner's proposal: the right hand side of
Kastner's Eq. $1''$ does not depend on $C$.
As  explained  in Section 3, the failure of the ABL rule for
hidden variables readings  is not surprising since the whole
concept of hidden variables is time asymmetric. (It might be
interesting to investigate the possibility of defining time-symmetric
hidden variables.) In any way, there are no hidden
variables  in the TSQT and, therefore, this failure does not represent a
problem.

It might be that Kastner and others have been mislead by the term
``element of reality''. The words suggest something ``ontological'',
but in the TSQT ``element of reality'' is a technical term which
describes a situation in which the outcome of a measurement is known
with certainty, see Section 4 and Vaidman (1996b). The only meaning of
an element of reality ``the particle is in box $A$'' quoted by Kastner
is that ``if searched in $A$ it has to be there with probability 1'',
nothing more.  After quoting this, Kastner writes (1999, p. 4[??]):
``This usage clearly implies that the properties of being in box $A$
or being in box $B$ are considered as possessed by the same pre- and
post-selected particle.'' The ``same'' means only that the two-state
vector at time $t$ is fixed. The counterfactual worlds corresponding
for ``being in $A$'' and ``being in $B$'' for the ``same'' particles
are different: in one world the particle is searched in $A$ and in
another it is searched in $B$.

\vskip 1.cm \noindent
 {\bf 8. Kastner's analysis   of the Sharp and
  Shanks proof.} 
\vskip 0.2cm
 
 Kastner  makes a distinction between two counterarguments which I
 presented in the two preprints (Vaidman, 1996a, 1997). From my
 point of view there is just one argument presented in different forms
 in the two preprints. The difference, as Kastner
 correctly noticed, is  that in Vaidman (1997) (which is  the revision
 of Vaidman 1996a) I do not focus on 
the possibility of a counterfactual with true antecedent. I still
think that this possibility is  a correct property of quantum counterfactuals.
However, I realized that many readers were confused by this point
and,  since it is not central, I decided that I can persuade
 better without emphasizing this property. 
 
 Kastner  writes that my counterarguments lead to what she
 calls ``non-counterfactual'' interpretation of the ABL rule ``which
 is not under dispute''. She then proceeds with the analysis of the
 S\&S argument focusing on ``a failure of cotenability between the
 background conditions, $S$, and the antecedent $P$''.  It seems to me
 that, this failure of cotenability is very similar to my argument
 against the proof of S\&S. They claimed that counterfactual
 interpretation of the ABL rule leads to predictions different from
 that of quantum theory. I claimed that their counterfactual
 interpretation has a logical error and therefore their proof is
 incorrect. Kastner shows that the left hand side of her Eq. 6 which
 is a part of the proof of S\&S is false. From false logical statement
 one cannot claim to calculate correctly probability for an outcome of
 a measurement, so the problem is not with the ABL formula, as S\&S
 claimed, but with the proof, as I claimed.
 
 However, Kastner, after writing that due to the failure of the
 cotenability condition ``it is clear that the counterfactual usage of
 the ABL rule fails'', continues with ``detailed
 description of the steps employed in the S\&S proof'', the ``proof
 that  the counterfactual interpretation of the ABL rule leads to
 predictions incompatible with quantum mechanics''. It seems to me
 that her detailed description leads to the same conclusion: the
 application of the counterfactual rule to the S\&S example is
 inconsistent -- exactly what I claimed -- and I do not understand the
 short paragraph in which she ``pinpoints the error in Vaidman's
 counterargument II''.  
 
 One of the difficulties in understanding Kastner's arguments in
 Section 2 and the apparent confusion in Section 3  follows from the
 usage of the concept of {\em mixture}. Kastner writes that that
 quantum ``measurement yields a mixture''.  The term ``mixture'' comes
 from statistics. It corresponds to a description with {\em density
   matrix}.  However, a result of a quantum measurement on an ensemble is
 described by a particular list of outcomes, not by a density matrix.
 The latter describes some statistical properties of this list, but it
 is not {\em the complete} description.  So, in order to proceed with
 Kastner's arguments I replace every word ``mixture'' by ``list of
 outcomes with statistical properties described by the mixture''.  In
 particular, the ``background condition'' should be changed from a
 ``mixture'' to a list of particular outcomes.
 
 The last paragraph of Section 2,  leaves me with several
 options for understanding Kastner's interpretation of the S\&S example.  The
 sentence ``In view of the existence of actual results at $t_2$, such
 results are an indelible part of the history of world {\it i} and
 cannot be disregarded.''  might suggest that the question is about a
 counterfactual world in which the measurement at time $t$ is
 performed, but, nevertheless the outcomes of the measurements at
 $t_2$ are as in the actual world.  (This is possible, although
 clearly a very improbable situation, since the ``mixture'' description
 of these outcomes is very different from the ``mixture'' expected by
 the laws of quantum theory for the situation with the
 intermediate measurement.) For {\em this} question, the counterfactual
 calculations of S\&S using the ABL formula are {\em correct} and the
 fact that it does not yield the value given by standard
 quantum-mechanical calculations,  Kastner's Eq. 10, is
 not a contradiction, because the latter corresponds to a different situation in
 which there is only pre-selection.
 
 Another possible reading of Kastner's paragraph is that in the
 counterfactual world with the intermediate measurement the outcomes
 of the measurement at time $t_2$, are different, but still the
 outcomes at $t_2$ in the actual world are relevant for calculating
 probabilities for the results of measurement at time $t$, ($t <
 t_2$).  One can imagine such a situation if there are hidden variables
 which control the outcomes of measurements beyond the standard
 quantum formalism. I have discussed this possibility in (iv) of
 Section 3. Indeed, in this case, the ABL formula yields  incorrect
 results, but this is not surprising since this situation is
 intrinsically time asymmetric: the actual and the counterfactual worlds
 coincide in the past, but not in the future, relative to time $t$.
 From private communications with S\&S I understand that the main goal
 of their paper was to show exactly this, i.e., that ``the ABL-rule
 did not have the implications for hidden variables interpretations of
 quantum mechanics that Albert et al. (1985) had
 claimed.''\footnote{This is a correct criticism which however,
   ``opens an open door''. The Letter of Albert et al.
   (1985) indeed leaves an impression that the authors make the
   discussion in the framework of the hidden variables theory.
   However, in their reply (Albert et al. 1986) to the criticism of Bub and Brown (1986)
   they clearly stated that they do not (or, at least, they do not now)
   think that the results of their Letter are applicable for
   hidden-variable theories.}  Although the conclusion is correct, I still think that the alleged proof in
 the paper of S\&S is flawed; I
 believe that I have proved it correctly in (iv) of Section 3. Careful
 reading of S\&S shows that they indeed focus on this limited issue.
 However, the title and the conclusions suggests criticism of the TSQT
  in much wider sense and lead their follower to attack all
 possible counterfactual interpretations in the framework of the TSQT.

\vskip 1.cm
 \noindent
{\bf 9. Kastner criticism of the time-symmetrized counterfactuals.~~ }
\vskip 0.2cm

Kastner again distinguishes between two, equivalent from my point of
view, definitions of time-symmetrized counterfactuals (TSC) given in
the two preprints (Vaidman 1996a, 1997).  I see the difference  only in
phrasing and the generality of their applications: Definition 2 is applicable
only for the ABL situation.

It seems that Kastner's criticism follows from a misunderstanding (which
I suspect came from her usage of the concept of mixture). Let me start
with the analysis of  Kastner's criticism of Definition 1. She
writes that in this definition I ignore the difference between
mixtures in actual and counterfactual worlds. The mixtures describe
the results of measurements at $t_1$ and $t_2$. But according to my
definition ``all measurements in a counterfactual world, excluding
measurements at $t$, have the same outcomes as in the actual world''.
Therefore, it is postulated by the definition that there is {\em no}
difference between the mixtures. There is nothing to ignore. Kastner's
quotation from my work about the difference which ``has no physical
meaning'' is related to the difference in the measurements (and
outcomes) performed at time $t$.  The outcomes of these measurements
depend on, but do not (by fiat) influence, the outcomes at $t_1$ and
$t_2$.

Kastner proceeds by alleged proof that in the example of the S\&S
setup there is no counterfactual world satisfying my definition. She
considers an ensemble of particles all pre-selected in a particular
spin state which are subject to spin measurement in another direction
at a later time. In the actual world no intermediate measurement is
performed. She claims to calculate the fraction of particular outcomes
of the second measurement in the ensemble. But she is not precise in
her claim that the expression she gets in Eq. 20,
$\cos^2(\theta_{ab}/2)$, is the fraction she wants to consider. Most
probably it is not equal to $k/N$ for any integer $k$, where $N$ is
the total number of particles in the ensemble. All that quantum theory
tells us is that in such an experiment it is likely to expect that the
fraction will be at the vicinity of $\cos^2(\theta_{ab}/2)$ of the
size of the order $1/\sqrt N$. But this fraction {\em can be any
  number} between 0 and 1 of the form $k/N$.  The calculation of this
fraction in the counterfactual world is meaningless: it has been {\em
  postulated} that it is equal to that of the actual world. Kastner's
Eq. 21 yields the quantum mechanical estimate of what to expect in an
alternative experiment. This calculation does not rule out identical
fractions and the even stronger requirement of identical outcomes in
the two experiments.
The counterfactual world I consider is possible. Therefore, the
definition is legitimate. The definition itself does not say for which
problems and in which situation it will be applied.  It is applied for
problems in which the pre- and post-selected states are fixed.
The question how often the post-selection succeeds is not under
discussion.

Other misunderstandings appear in Kastner's criticism of Definition
2. First, she augments the two-state vector notation with a
specification which observable has been measured at time $t$. This is
against the whole idea of the TSQT. The two-state vector is a complete
description at time $t$ in the sense that it yields probabilities for
all possible measurements (and the weak values for weak measurements)
at time $t$. The two-state vector is specified by the results of
measurement at times {\em different} than $t$: the measurement at $t$ does
not have direct influence on the two-state vector at $t$. See Vaidman
(1998b) for careful review of this concept. 

Kastner, instead, makes her own definition of ``time-symmetrically
fixed''. Although she writes that it is ``in the sense of Definition
2'', her definition has nothing in common with my proposal. In my
proposal there is no question: will the system have the same
two-state vector? The two-state vector is given by fiat and this {\em is}
the ``time-symmetrical fixing''.
   It seems that behind Kastner's definition there is an idea of some
kind of hidden variable: she discusses systems which ``would still
have the same two-state vector'' even so some different measurement
are performed at $t$. To have the same two-state vector is, in the
current context, to have the same outcome of a measurement at $t_2$
in a situation in which {\em a priory} it is not certain due  to
quantum-mechanical laws. Standard quantum theory does not ensure {\em
  the same} outcome at $t_2$ even if the same measurement is performed at $t$
and the situation is different only because of a change in some
unrelated variable. The example which  Kastner considers  demonstrates
how she distorts  Definition 2. The meaning of ``the results of
measurements performed on the system at times 
$t_1$ and $t_2$ are fixed'' is that  the outcomes of the
measurements at  $t_1$ and at $t_2$ in actual and counterfactual worlds are the
same. In Figure 4 it is not true. Only the outcomes of the measurement
at $t_1$ are the same.

\vskip 1.cm \noindent {\bf 10. Conclusions} 
\vskip 0.2cm

The definition of counterfactuals in quantum theory which I propose is
very simple-minded. It seems to me that if one reads it as it is,
without trying to find something beyond it,\footnote{Apart from
  ``redefinition'' of Definition 2,  Kastner uses  the word
  ``ontological'' in her paper. The TSQT
  does not make ontological claims. It is more a novel {\em formalism}
  than an {\em interpretation}.}, then it is complete and unambiguous.
I believe that the definition is helpful in resolving some
controversies about quantum counterfactuals (see my attempts in this
direction in Vaidman (1998c, 1998d)). 

In principle, the counterfactual
statement such as definition 2 is testable in a laboratory by creating
a large ensemble of systems with measurements at $t_1$, $t$, and
$t_2$; choosing the sub-ensemble (the pre- and post-selected ensemble)
with fixed outcomes at $t_1$ and $t_2$; choosing (out of this
sub-ensemble) the sub-ensemble with a particular measurement at $t$;
and finally by
making a statistical analysis of the outcomes of the measurement at
$t$ on this sub-ensemble.
Because it is testable, philosophers might be reluctant to
consider the construct which I define as counterfactual, in spite of
the fact that formally it corresponds to the counterfactual (i) of Section
3.  This is a semantic
issue. I distinguish in Section 4 between situations in which only
single statement of the form of Definition 2 is considered and
situations in which several such statements, for different variables
are considered all related to a single system. Since quantum theory
does not allow  simultaneous measurements of certain variables, in the
latter situation the set of statements is, in fact, not testable.

 I think that the most
convincing example that the term counterfactual is appropriate and that the whole
construction is useful is the example with three boxes (Vaidman 1996b).
A single quantum system is weakly coupled to another system between two
measurements at $t_1$ and $t_2$. For this system several counterfactual (in my sense)
statements of the form of Definition 2 about intermediate measurements
on this system with fixed outcomes of the measurements at $t_1$ and
$t_2$ are true. This statements seems to be clearly counterfactual. I
know that  the
measurements have not been performed because  in the actual world the only
interaction at time $t$ is a weak coupling to another system.  Moreover, since there is only one
system and the measurements are incompatible, I know that they all
could not have been performed. Finally, the knowledge of these counterfactual
statements  permits me to calculate the expected influence of the weak
coupling on the other system which takes place in the actual world.

Kastner concludes that ``counterfactual interpretation of te ABL rule
is not valid in general''. I have showed that this conclusion follows
from Kastner's particular reading of the word ``counterfactual''.  I
find that by and large  Kastner  supports my claim that counterfactuals in
a sense which  differs  from mine are inconsistent in the
framework of the TSQT. She rejects my approach saying that this is a
``non-counterfactual'' reading.  I disagree about this semantic issue.
More importantly, I disagree with Kastner's claim that in several works
in the framework of the TSQT the ``bona fide'' counterfactual reading of
the ABL rule has been used (from which it follows that, since this
reading is inconsistent, the results of these works, various
``curious'' quantum effects, are wrong). This claim, however was not
proved and only stated in Kastner's paper. Indeed, essentially only time-asymmetric
examples were analyzed in her paper.

I have showed that Kastner's criticism of my definitions of TSC is
unfounded. In her discussion of Definition 1 she erroneously considers
``improbable'' as ``impossible'' neglecting the fact that the TSC are
applicable to pre- and post-selected situations which are usually
``rare quantum events''. The Definition 2 she interprets in a
particular sense. In this sense it is  indeed ``has no clear physical
meaning''. I suspect that   she rejects the literal interpretation of
Definition 2, the one which I  adopt, because
she views it as non-counterfactual.

Kastner finds ``an interesting special case in which the ABL rule may
be correctly used in a counterfactual sense'', the one which
corresponds to {\em consistent histories} (Griffiths, 1984). In the
ABL case this corresponds to measurements of an observable for which
either a pre-selected or post-selected state is an eigenstate and,
therefore, the outcome is certain. This case seems to me extremely
limited and very uninteresting. Even the case in which the outcome is
certain, but it is not an eigenvalue of either the pre-selected or
post-selected state as in the there-box example, does not fall into
this category. I do not find fruitful a concept which is applicable to
a very limited class of situations, especially if one can consider a
similar alternative concept which is applicable for all situations.

Many quantum mechanical effects are dramatically different from
phenomena which can be explained classically. Language and philosophy
which were developed during the time that no one suspected quantum
phenomena, have significant difficulties in defining and explaining
quantum reality. This seems to be the reason for numerous
controversies in this field. I believe that discussing and resolving
these controversies is of crucial importance for understanding our
world.

It is a pleasure to thank Yakir Aharonov, David Albert, Avshalom
Elitzur, Lior Goldenberg, Yoav Ben-Dov, Igal Kvart,  Abner Shimony and
Stephen Wiesner for helpful discussions and Willy De Baere, Jonathan
Bennett, Ruth Kastner, Philip Pearle, Niall
Shanks, and David Sharp for useful correspondence. 
 The research was supported in part by grant
471/98 of the Basic Research Foundation (administered by the Israel
Academy of Sciences and Humanities).

\vskip .8cm
\vfill
\break

 \centerline{\bf REFERENCES}
\vskip .15cm

\vskip .13cm \noindent 
Aharonov, Y. and Albert, D. (1987), 
`The Issue of Retrodiction in Bohm's Theory',
in B.J. Hiley and F.D. Peat (eds.) {\em Quantum Implications}. New
York: Routledge \& Kegan Paul, pp.224-226.

\vskip .13cm \noindent 
Aharonov, Y.  Albert, D. and Vaidman, L (1986),
`Measurement Process in Relativistic Quantum Theory',
{\it Physical  Review} {\bf  D 34}, 1805-1813.
 
\vskip .13cm \noindent 
Aharonov, Y.,  Bergmann,  P.G., and  Lebowitz, J.L. (1964),
`Time Symmetry in the Quantum Process of Measurement',
 {\em Physical Review}  {\bf 134B}: 1410-1416.

\vskip .13cm \noindent 
Aharonov, Y. and Vaidman, L. (1990),
 `Properties of a Quantum System
During the Time Interval Between Two Measurements',
{\em Physical Review}  {\bf A 41}, 11-20.

\vskip .13cm \noindent 
Aharonov, Y. and Vaidman, L. (1991),
`Complete Description of a Quantum System at a Given Time',
{\em Journal of  Physics} {\bf  A 24}, 2315-2328.

\vskip .13cm \noindent 
Albert, D. Z., Aharonov, Y., and D'Amato, S. (1985), 
 `A Curious New Statistical Prediction of Quantum Theory', {\it Physical Review Letters} {\bf 54}, 5-7.

\vskip .13cm \noindent 
Albert, D.,  Aharonov, Y., and  D'Amato, S. (1986),
`Comment on `Curious Properties of Quantum Ensembles which have been Both
  Preselected and Post-Selected',
{\em Physical Review Letters} {\bf 56}, 2427.

\vskip .13cm \noindent 
Bedford, D. and Stapp, H.P. (1995),
`Bell's Theorem in an Indeterministic Universe',
{\em Synthese} {\bf 102}, 139-164.

\vskip .13cm \noindent 
Bennett, J. (1984),
`Counterfactuals and Temporal Direction',
{\em Philosophical Review} {\bf 93}, 57-91.

\vskip .13cm \noindent 
 Bub, J. and Brown, H. (1986), 
`Curious Properties of Quantum Ensembles which have been Both
  Preselected and Post--Selected',
{\em Physical Review Letters} {\bf 56}, 2337-2340.

\vskip .13cm \noindent 
Cohen,  O. (1995), 
`Pre- and Post-Selected Quantum Systems, Counterfactual Measurements,
and Consistent Histories',
 {\em Physical Review} {\bf  A 51}, 4373-4380. 

\vskip .13cm \noindent 
 Cohen, O. and Hiley, B.J. (1996), 
`Elements of Reality, Lorentz Invariance and the Product Rule',
 {\em Foundations of Physics} {\bf 26}, 1-15. 

\vskip .13cm \noindent 
Ghirardi, G. C. and Grassi R. (1994),
`Outcome Predictions and Property Attribution: the EPR Argument
Reconsidered',
{\em Studies in History and Philosophy of Science} {\bf 25}, 397-423.

\vskip .13cm \noindent 
Griffiths, R. B. (1984), `Consistent Histories
and the Interpretation of Quantum Mechanics,' 
{\it Journal of Statistical 
Physics} {\bf  36}, 4373.

\vskip .13cm \noindent 
Kastner, R. E. (1999),
`Time-Symmetrized Quantum Theory, Counterfactuals,
and `Advanced Action',
{\em Studies in History and Philosophy of Science},  this issue, quant-ph/9806002.

\vskip .13cm \noindent 
Lewis, D. (1973),
{\em Counterfactuals},
Oxford: Blackwell Press.

\vskip .13cm \noindent 
 Lewis, D. (1986),
 `Counterfactual Dependence and
  Time's Arrow' reprinted from {\em Nous 13}: 455-476 (1979) and {\em
    Postscripts to `Counterfactual ..'} in Lewis, D. {\em
    Philosophical Papers Vol.II}, Oxford: Oxford University Press,
  pp.32-64.

\vskip .13cm \noindent 
Mermin, N.D. (1989),
`Can You Help Your Team Tonight by Watching on TV? More Experimental
Metaphysics from Einstein, Podolsky, and Rosen',
in J.T. Cushing and E. McMullin (eds.) {\em Philosophical Consequences
  of Quantum Theory: Reflections on Bell's Theorem}. Notre Dame:
University of Notre Dame Press, pp.38-59.

\vskip .13cm \noindent 
Miller, D. J. (1996),
`Realism and Time Symmetry in Quantum Mechanics',
{\em Physical Letters} {\bf  A 222}, 31-36.

\vskip .13cm \noindent 
Penrose, R. (1994),
{\em Shadows of the Mind}.
 Oxford: Oxford University Press.

\vskip .13cm \noindent 
Peres, A. (1978),
`Unperformed Experiments Have no Results',
{\em American Journal of Physics } {\bf 46}, 745-747.

\vskip .13cm \noindent 
Peres, A. (1993),
{\em Quantum Theory: Concepts and Methods}.
Dordrecht: Kluwer Academic Publisher.

\vskip .13cm \noindent 
Price H. (1996),
{\em Time's Arrow \& Archimedes' Point}. New York: Oxford University Press.

\vskip .13cm \noindent 
Redhead, M. (1987),
{\em Incompleteness, Nonlocality, and Realism: a Prolegomenon to the
  Philosophy  of Quantum Mechanics}. New York: Oxford University Press.

\vskip .13cm \noindent 
Sharp, W.D. and Shanks, N. (1993),
`The Rise and Fall of Time-Symmetrized Quantum Mechanics',
{\em Philosophy of Science} {\bf  60}, 488-499.

\vskip .13cm \noindent 
 Skyrms, B. (1982),
 `Counterfactual Definiteness and Local
 Causation', {\em Philosophy of Science} {\bf  49}, 43-50.

\vskip .13cm \noindent 
 Vaidman, L. (1993),
`Lorentz-Invariant `Elements of Reality' and the Joint
Measurability of Commuting Observables',
{\em  Physical Review Letters} {\bf  70}, 3369-3372.

\vskip .13cm \noindent 
Vaidman, L. (1996a), `Defending Time-Symmetrized Quantum 
Theory,' e-print 
quant-ph/9609007.

\vskip .13cm \noindent 
Vaidman, L. (1996b),
`Weak-Measurement Elements of Reality',
 {\em Foundations of Physics} {\bf  26}, 895-906.

\vskip .13cm \noindent 
Vaidman, L. (1997),  `Time-Symmetrized Counterfactuals in Quantum 
Theory', Tel-Aviv University preprint TAUP 2459-97,  
quant-ph/9807075.

\vskip .13cm \noindent 
Vaidman, L. (1998a),
`Validity of the Aharonov-Bergmann-Lebowitz rule',
{\it Physical Review} {\bf A 57}, 2251-2253.

\vskip .13cm \noindent 
Vaidman, L. (1998b),  Time-Symmetrized Quantum Theory,
{\it Fortschritte Der Physik}, {\bf 46}.

\vskip .13cm \noindent 
Vaidman, L. (1998c),  `Time-Symmetrized Counterfactuals in Quantum 
Theory', e-print
quant-ph/9807042.

\vskip .13cm \noindent
 Vaidman, L. (1998d), Variations on the Theme of
the Greenberger-Horne-Zeilinger Proof', {\em Foundations of Physics},
to be published, e-print
quant-ph/9808022.

\end{document}